# The Abundance of Low Energy Cosmic Ray Boron and Nitrogen Nuclei Measured at Voyager 1 Beyond the Heliopause – Where Do They All Come From? – An Interpretation Using a Leaky Box Galactic Propagation Model

**W.R. Webber**

New Mexico State University, Astronomy Department, Las Cruces, NM 88003, USA

**ABSTRACT**

The intensities and spectra of the secondary Boron and Nitrogen nuclei from a three year study of Voyager data beyond the heliopause are interpreted using a Leaky Box model for galactic propagation. For B, which is a purely secondary nucleus, in the energy range ~10-40 MeV/nuc, the path length in the galaxy required to produce the observations is at least 10 g/cm$^2$. At energies between 40-120 MeV/nuc the path length deduced from the Boron and Nitrogen nuclei measurements is smaller. These higher energy measurements above 40 MeV/nuc are, in fact, compatible with the path length that has been used to propagate primary H and He nuclei, namely $\lambda = 23.3 \, \beta \, P^{-0.47}$ (g/cm$^2$) above a value of P between 0.316–1.0 GV and $\lambda$=const $\beta^{3/2}$ (g/cm$^2$) below these rigidities. At 40 MeV/nuc this path length would be between 4-10 g/cm$^2$. Uncertainties in the Voyager data itself as well as the cross sections for production of these secondaries at these low energies are an important limitation on the estimates of the amount of matter traversed. We also attempt a fit to the B observations at low energies by considering a "source" component of B generated in ~1.0 g/cm$^2$ of matter near the sources but after acceleration. This would be equivalent to a Nested LBM. In this model the V1 data below ~40 MeV/nuc is now well fit. For Nitrogen, which is dominated by a source component at these low energies, a fit to the data between 10-130 MeV/nuc can be obtained with a N/O source ratio = 6.3 $\pm$ 1.0% and a path length compatible with the H and He propagation as noted above. This Voyager observation confirms earlier measurements of a low N abundance in the cosmic ray source relative to what is found for solar abundances, for example, where the N/O ratio is found to be ~12%.

## Introduction

Cosmic Ray nuclei from H to Fe have now been measured for a 3 year time period beyond the heliopause at Voyager 1 (Cummings, et al., 2015). There now are reasonable statistics to measure and compare with propagation models for the primary (source) nuclei H, He, C, O, Ne, Mg, Si and Fe (Webber, 2015).

The measurements of the nuclei B and N reported in Cummings, et al., 2015, and used here are the only secondary nuclei from Voyager with sufficient statistical accuracy at present to bring to bear on some of the questions arising from a comparison of observations and predictions of nuclei propagation in the galaxy. In this paper we will show that the measured intensity of B below about 40 MeV/nuc requires a mean path length of at least 10 g/cm$^2$ in a Leaky Box Model (LBM), for its production during interstellar propagation, whereas at energies >40 MeV/nuc a path length, $\lambda = \sim\beta^{3/2}$ for $P_0$ between 0.316-1.0 GV and $\lambda = 22.3 \beta P^{-0.47}$ g/cm$^2$ above $P_0$, the same as used to explain the observed spectra of primary nuclei (Webber, 2015), will explain the B data. This will provide a path length ranging from ~4 g/cm$^2$ at 40 MeV/nuc to ~10 g/cm$^2$ at 100 MeV/nuc.

For N nuclei, the observed abundance is dominated at the lowest energies by a source abundance which is ~6% of that of O and the overall N abundance is therefore a composite between this source abundance and secondary production in the galaxy.

## The Data

The V1 data for B and N nuclei are shown in Figures 1 and 2 respectively (see Cummings, et al., 2015). The calculations using the Leaky Box propagation model are shown as solid black lines in Figures 1 and 2.

These propagation calculations are similar to those described in Webber and Higbie, 2009, and updated in Webber, 2015. The mean path length in the galaxy is described above $P_0$ by the expression $\lambda = 22.3 \beta P^{-0.47}$ in g/cm$^2$ in the latest calculation. These values for $\lambda$ are based on new measurements of the B/C ratio at both low and high energies (Webber, 2015; Webber and Villa 2016). The parameter $P_0$ describes the rigidity at which the cosmic ray diffusion coefficient in the galaxy changes its rigidity dependence by ~1.0 power from being ~$P^{0.47}$ at high rigidities to being $\sim P^{-0.53}$ at low rigidities. This changes the dependence of the path length $\lambda$ to $\lambda$ = const. $\cdot \beta^{3/2}$ at lower rigidities. This change of 1.0 power in the rigidity dependence of the diffusion coefficient is determined from fits to the Voyager 1 measured electron spectrum at low energies and the PAMELA electron spectrum measured at higher energies up to 10 GeV/nuc where the solar modulation is small (Webber and Higbie, 2013).

The parameter, $P_0$, in effect, thus determines the amount of interstellar material, $\lambda$ in g/cm$^2$, that cosmic rays traverse at low energies (1.0 GV $\sim$120 MeV/nuc for A/Z = 2.0 nuclei). This dependence is illustrated in Figure 3 which shows the values of $\lambda$ in g/cm$^2$ as a function of P and also E for various assumptions for the value of $P_0$. For a value of $P_0$=1.0 GV the path length = 1.5 g/cm$^2$ at an energy of 12 MeV/nuc. If $P_0$=0.316 GV the path length = 7.6 g/cm$^2$ at this same energy.

For Boron, in Figure 1, where there is no source component, the Voyager observations, with uncertainties $\pm$ 15-25% at the lowest energies (both experimental and cross section uncertainties for each data point) lie generally above the predictions below $\sim$40 MeV/nuc even for a value of the path length as large as 10 g/cm$^2$. The difference between the LBM predictions and the measurements is within the Voyager statistical errors and cross section errors above about 40 MeV/nuc.

For Nitrogen in Figure 2 the source component dominates the N abundance at the lowest energies (see below). A combination of a source component $^{14}$N = 6.3 $\pm$ 1.0% of the abundance of $^{16}$O and a value of $P_0$ between 0.316 to 1.0 GV with $\lambda \sim \beta^{3/2}$ below $P_0$ for the propagation then fits the Nitrogen data between 10-130 MeV/nuc.

## The Calculation of B and N Production in Galactic Cosmic Rays

The calculations for B and N are made using the same set of propagation parameters that have been used to calculate the abundances of the primary nuclei H through Fe in the Leaky Box propagation program that we use (Webber and Higbie, 2009; Webber, 2015). The "standard" calculations use a mean path length $\lambda$ = 22.3 $\beta P^{-0.47}$ above $P_0$ in a pure Leaky Box Model where the path length distribution at each energy is exponential. Below $P_0$ the rigidity dependence of the diffusion coefficient is assumed to change from being $\sim P^{0.47}$ at higher rigidities to a dependence ~$P^{-0.53}$ at lower rigidities. The calculations in Figures 1 and 2 are shown for values of $P_0$=0.316 and 1.0 GV. The secondary production is larger at the lowest energies by a factor ~2 for a value of $P_0$=0.316 GV because of the larger value of the mean path length (see Figure 3).

If the path length would remain constant below 1.0 GV at a value 10.7 g/cm$^2$ the resulting B production is also shown in Figure 1 as a solid line.

The cross sections into B and N from $B^{11}$, $C^{12}$, $N^{14}$, $N^{15}$ and $O^{16}$, which provide ~75% of all B production and nearly 50% of all N production and are used in the calculation, are shown in Table 1. The stripping cross sections where a proton or neutron is striped off the progenitor nucleus are an important factor in the cross sections below ~100 MeV/nuc and are not well measured (see Maurin, Putze and Derome, 2011; Coste, Derome, Maurin and Putze, 2012). The calculations of Silberberg and Tsao, 1973, are useful in the estimates shown in the table, which have uncertainties of as large as $\pm$ 20% below ~100 MeV/nuc, the same order or larger than the current Voyager data statistical uncertainties at these energies.

**Discussion**

The abundance of B measured by Voyager below ~40 MeV/nuc and described in this paper requires that the path length for production of the secondary B nuclei to be at least 10 g/cm$^2$ in a simple Leaky Box Model. This large path length is not compatible with a change in the rigidity dependence of the propagation (diffusion coefficient) from ~$P^{0.47}$ above $P_0$ to a dependence ~$P^{-0.53}$ below $P_0$, which is what is required to fit the V1 electrons (Webber and Higbie, 2013). This is true even for values of $P_0$ less than 0.316 GV in which case the high rigidity dependence extends down to $P_0$ (see Figure 3 and Figure 1 where the calculations for $P_0$=0.316, and for $\lambda$= 10.7 g/cm$^2$ are shown).

We note that the recent calculations from GALPROP (Cummings, et al., 2015) shown in Figure 1 also under-predict the B abundance at the lower energies. These calculation quite closely follow the curve for $P_0$=0.316 GV from our calculations.

We therefore ask the question, could the observed high B abundances at low energies be produced if we assume that B has a "source" abundance to begin with?  In Figure 4 we show a calculation of the B abundance for a "source" abundance of B which ~5% of the source abundance of C.  To produce this B abundance requires that cosmic rays travel through ~1.0 g/cm$^2$ of matter near the source, but after acceleration as in Nested LBM Models (Cowsik and Wilson, 1973; Cowsik, Burch and Madziwa-Nussinov, 2014).  This is in addition to the B production during the galactic propagation which therefore needs to be reduced by ~15% to match the measured B/C ratio (see Webber and Villa, 2016).  In this case, the overall calculation of the B abundance proceeds in two stages.  In the first stage the cosmic rays traverse ~1 g/cm$^2$ thus producing a B component which is ~5% of C.  This fraction is almost independent of energy above ~50 MeV/nuc and acts almost like a pure secondary component similar to the secondary N component described below.  The second step in the calculation is to propagate this modified source distribution.  The combination of the two stages should fit the observed B/C ratio.

Such a source component could arise if some B would be produced after the acceleration, but in material surrounding the source.  This is the concept of a Nested LBM.  In our picture the amount of material near the sources is a small fraction (~1 g/cm$^2$) of the overall material traversal in interstellar space.  In the picture by Cowsik, Burch and Madziwa-Nussinov, 2014, used to explain high energy positron production, the amount of matter at the source is large and only ~1.7 g/cm$^2$ is traversed in interstellar space.

In the calculation shown in Figure 2 we have assumed that the fraction $^{14}$N/O =6.30% and the fraction $^{15}$N/O = 0.35%.  For N nuclei this source abundance is the dominant component of the observed N abundance below about 60 MeV.  This can be seen in Figure 2 where we show calculations of the N abundance in the Leaky Box Model for source abundances of $^{14}$N of 0% and 3% of O.  The 0% curve would correspond to the pure secondary production of N in the galaxy.  This secondary fraction decreases from ~60% of the total N abundance observed at 100 MeV/nuc to less than 20% at ~10 MeV/nuc.  Calculations show that, if the source abundance of $^{14}$N/O is changed by $\pm$ 0.010, the total abundance of N is changed by $\pm$ 18% at 10 MeV/nuc and ~15% at 80 MeV/nuc (these are the measured quantities).  From these differences in the calculated N abundance, which are very similar to the uncertainties in both the Voyager measurements and the production cross sections into N, we estimate that $^{14}$N = 0.063 $\pm$ 0.010 of O. This source ratio can be improved somewhat with a longer integration time beyond the heliopause but is limited by the cross section errors.

In the case of added N production near the source as in a Nested LBM, for an N production in 1.0 g/cm$^2$ of material near the source, the "initial" source abundance 6.3% of $^{14}$N relative to Oxygen would need to be decreased by ~2.4% from 6.3 to~3.9% to match the observed N abundances.

## Summary and Conclusions

In this paper we compare the abundances of the secondary charges Boron and Nitrogen, measured by V1 over a 3 year time period beyond the heliopause, with calculations using a Leaky Box Model for galactic propagation. We find for B: The measured B abundance below ~40 MeV/nuc requires a large amount of interstellar matter traversal, ~10 g/cm$^2$ or larger. At higher energies, from 40 to 120 MeV/nuc the interstellar matter traversal required to produce the observed intensity of B is consistent with that for a matter traversal defined by values of $P_0$ between 0.316 and1.0 GV and a value $\lambda$=const $\beta^{3/2}$ g/cm$^2$ below $P_0$. This results in a matter transversal of between ~1.5 and 7.6 g/cm$^2$ at 12 MeV/nuc increasing to ~10 g/cm$^2$ at 120 MeV/nuc (see Figure 3). The uncertainties of the cross sections into B below ~120 MeV/nuc and the uncertainties in the Voyager data itself limit the accuracy of this measurement at the lowest energies.

The addition of a "source" component of B which is equivalent to the production of B in ~1.0 g/cm$^2$ of material traversed by cosmic rays near the source, but after acceleration (essentially a Nested LBM) will also produce an abundance of B that is consistent with the abundance measured by V1 but slightly larger than the value of $\lambda$=10.7 g/cm$^2$ of interstellar matter obtained from the simple LBM (see Figure 4).

We find for N: The measured abundance between 10-130 MeV/nuc can be fit accurately using the normal propagation parameters with $P_0$ between 0.316 and 1.0 GV and $\lambda$ ~$\beta^{3/2}$ below $P_0$. This leads to a source abundance $^{14}$N = 6.3% $\pm$1.0% of O. This source abundance represents about 80% of the total abundance of N that is measured at the lower energies, i.e., it dominates over secondary production which is only ~20% of the total N abundance at 10 MeV/nuc. Thus the abundance of N that is measured at these very low energies may be used to determine the source abundance of $^{14}$N with an accuracy which is limited mainly by the statistical accuracy of the Voyager data and should improve with time.

Earlier measurements of the N abundance by Voyager in the inner heliosphere and subject to solar modulation effects have determined a $^{14}$N source abundance of between ~4-6% of O (Webber, et al., 1996). Thus our new measurements confirm the earlier measurements and indicate a $^{14}$N abundance that is between 0.33 – 0.50 times a solar abundance of 12% of O.

The addition of a "source" component of N equivalent to the production of N in 1.0 g/cm$^2$ of material traversed by cosmic rays near the sources would reduce the estimated true source abundance to 3.9% of O as compared with 6.3% in a pure LBM.

**Acknowledgements:** The author appreciates the support of JPL. Paul Higbie made many valuable contributions to the application of the computer programs used herein.

| | | | | | | | | **E (MeV/nuc)** | | | | | | | |
|---|---|---|---|---|---|---|---|---|---|---|---|---|---|---|---|
| | **10** | **20** | **40** | **60** | **80** | **100** | **150** | **200** | **300** | **400** | **600** | **800** | **1000** | **2000** | **5000** |
| 11B | 46 | 66 | 86 | 86 | 76 | 62 | 50 | 445 | 445 | 445 | 445 | 445 | 490 | 430 | 430 |
| 12C | 122 | 152 | 146 | 138 | 128 | 115 | 101 | 86 | 82 | 82 | 82 | 82 | 81 | 78 | 75 |
| 14N | 21 | 37 | 46 | 51 | 47 | 41 | 39 | 39 | 39 | 39 | 39 | 39 | 39 | 38 | 31 |
| 15N | 45 | 45 | 89 | 49 | 48 | 45 | 42 | 42 | 42 | 42 | 42 | 42 | 41 | 41 | 40 |
| 16O | 16 | 36 | 39 | 39 | 39 | 39 | 40 | 40 | 40 | 40 | 40 | 40 | 40 | 39 | 38 |

**TABLE 1**
**Cross Sections in mb**
**to $^{10}B + ^{11}B$**

# REFERENCES

Coste, B., Derome, L., Maurin, D. and Putze, A., 2012, A&A, <u>539</u>, A88

Cowsik, R. and Wilson, L.W., 1973, Proc. 13[th] ICRC, Denver, 1,5

Cowsik, R., Burch, B. and Madziwa-Nussinov, T., 2014, Ap.J., <u>786</u>, 7

Cummings, A.C., Stone, E.C. and Lal, N., et al., 2015, Proc. ICRC,

Maurin, D., Putze, A. and Derome, L., 2010, A&A, <u>516</u>, 67

Silberberg, R. and Tsao, C.H., 1973, Ap.J., <u>25</u>, 315

Webber, W.R. 1993, Ap.J., <u>402</u>, 188-194

Webber, W.R., Lukasiak, A. and McDonald, F.B., et al., 1996, Ap.J., <u>457</u>, 435-439

Webber, W.R. and Higbie, P.R., 2009, JGR, <u>114</u>, 2103-2109

Webber, W.R. and Higbie, P.R., 2013, http://arXiv.org/abs/1308.6598

Webber, W.R., 2015, http://arXiv.org/abs/1508.01542

Webber, W.R. and Villa, T.L., 2016, http://arXiv.org/abs/ (unpublished)

**FIGURE CAPTIONS**

**Figure 1:** Voyager 1 measurements of B nuclei intensities beyond the heliopause. Statistical uncertainties are shown. Solid lines show the calculated intensities for a LBM with a mean path length, $\lambda = 22.3 \; \beta \; P^{-0.47}$ g/cm$^2$ and values of $P_0 = 0.316$ and 1.0 GV and with $\lambda \sim \beta^{3/2}$ below $P_0$, and also for a value $\lambda = 10.7$ g/cm$^2$, constant with rigidity below 1.0 GV. The GALPROP calculated values come from Cummings, et al., 2015.

**Figure 2:** Same as Figure 1 but for N nuclei. The calculated N abundances for a $^{14}$N source equal to 0% and 3% of $^{16}$O are shown as dashed lines. Note that the line for 0% represents the production of secondary nuclei only which equals ~20% of the observed abundance of N at 10 MeV/nuc so, in fact, at this energy most of the measured N abundance is from the source component.

**Figure 3:** Mean path length, $\lambda$ in g/cm$^2$ used in LBM calculations for B and N nuclei.

**Figure 4:** Similar to Figure 1 but shows the B production calculated using a "source" component equivalent to 0.5 g/cm$^2$ of material traversed near the source, in addition to the normal LBM propagation through the galaxy with $P_0 = 1.0$ GV and the galactic path length reduced by ~15%.

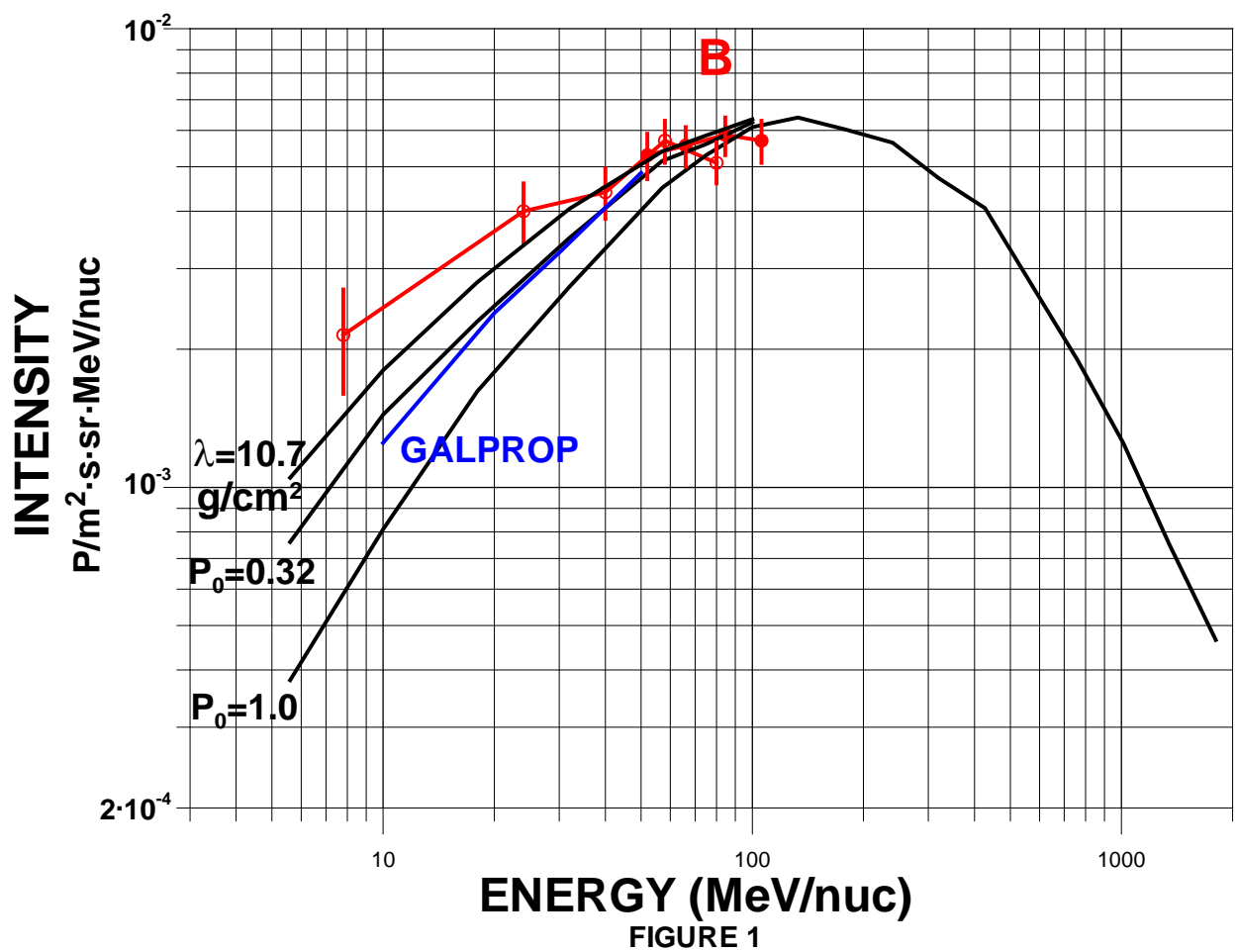

FIGURE 1

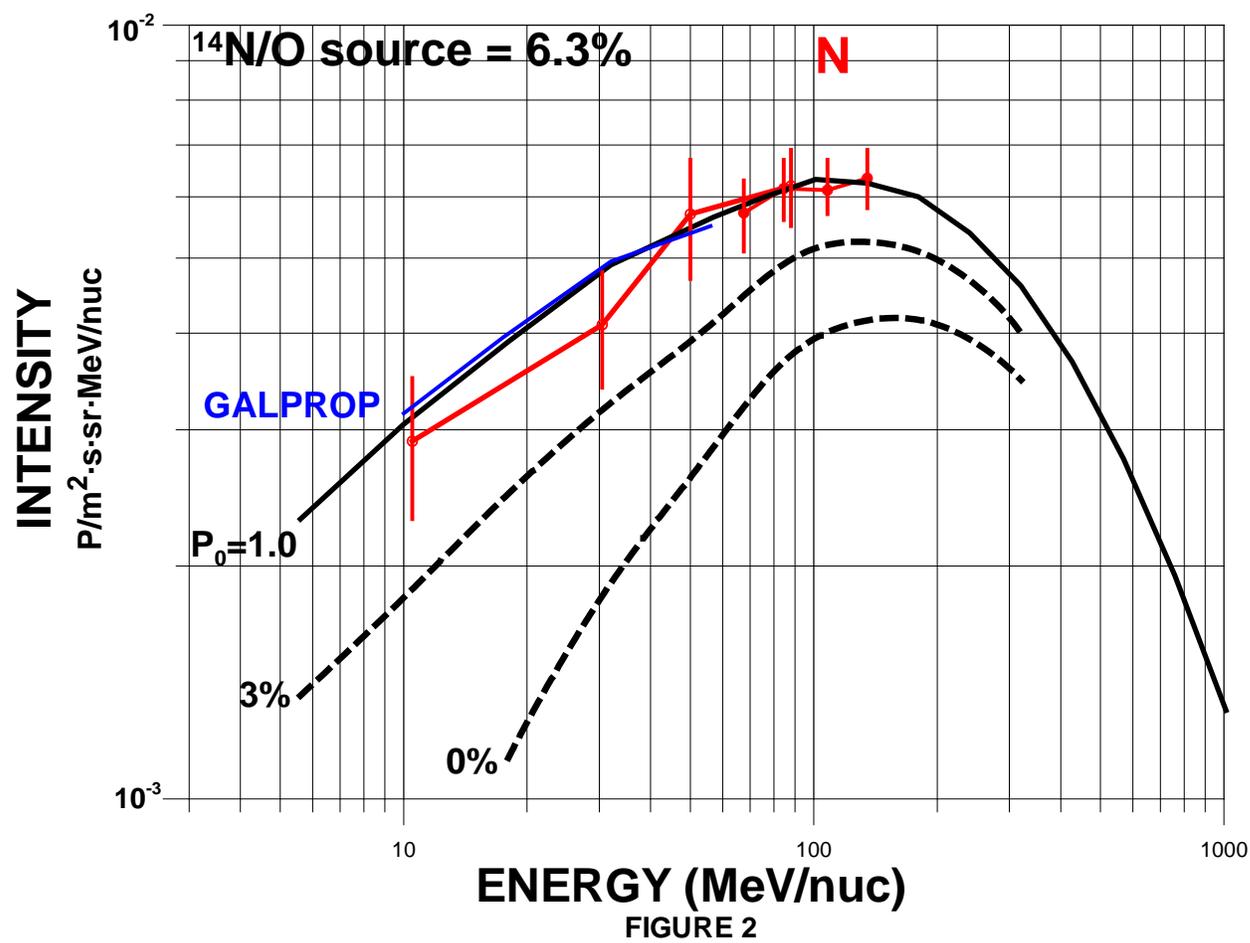

FIGURE 2

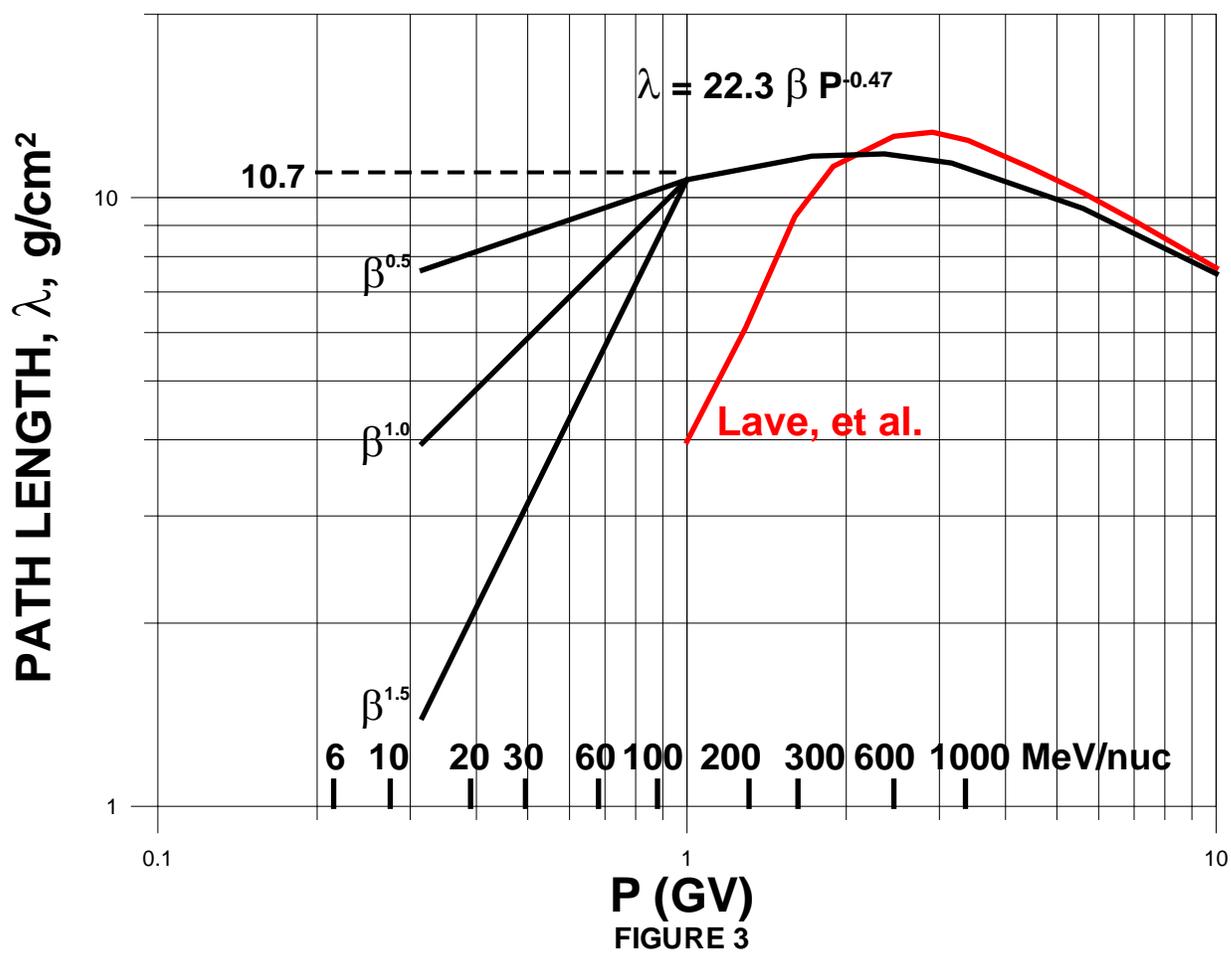

FIGURE 3

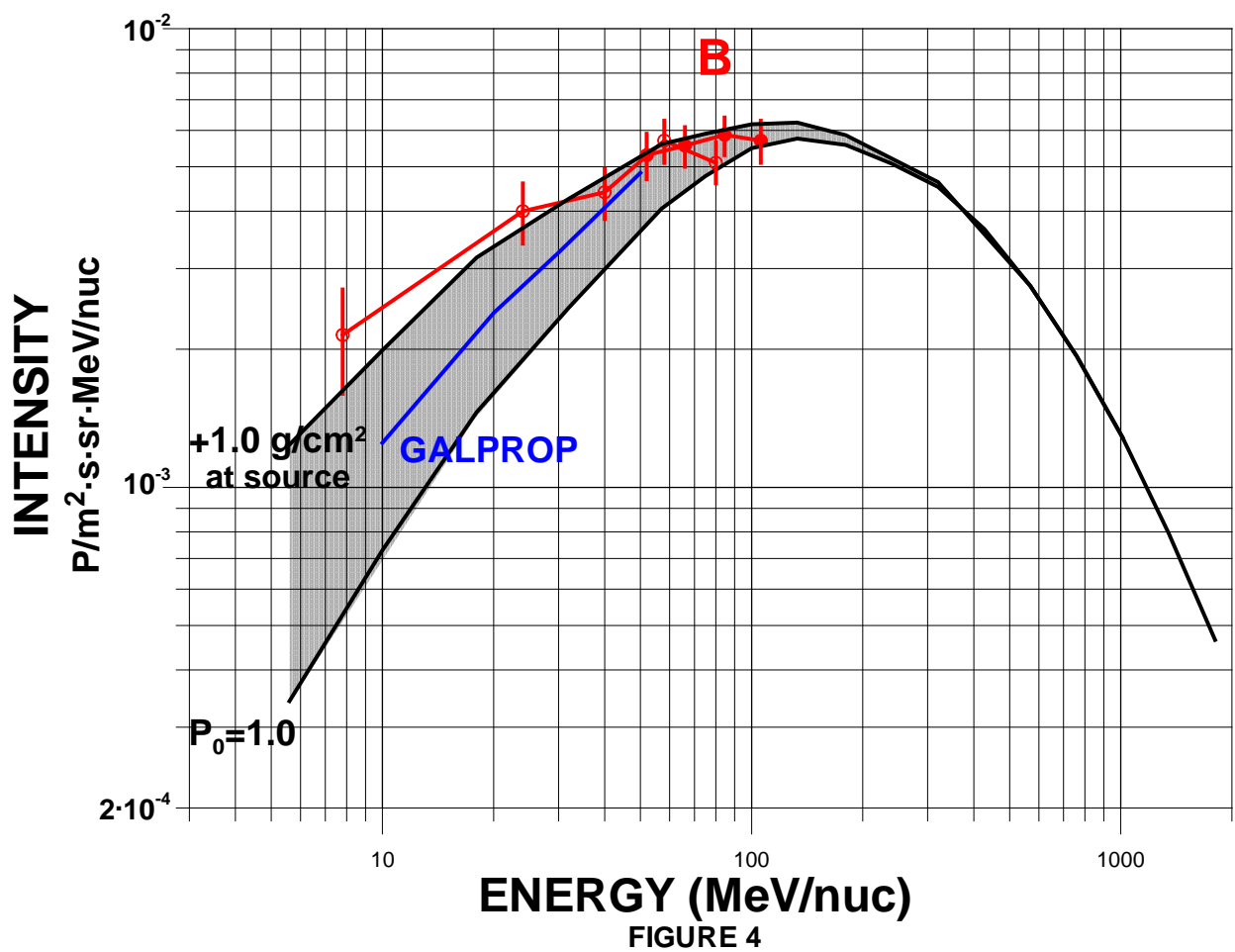

INTENSITY
P/m²·s·sr·MeV/nuc

**B**

+1.0 g/cm²
at source

GALPROP

P₀=1.0

**ENERGY (MeV/nuc)**

FIGURE 4